\pdfoutput=1
\documentclass[aps,prb,10pt,twocolumn,showpacs,superscriptaddress,nofootinbib]{revtex4-1}
\usepackage{amsmath}
\usepackage{latexsym}
\usepackage{amssymb}
\usepackage{bm}
\usepackage{color}
\usepackage{amsmath}
\usepackage[usenames,dvipsnames,svgnames]{pstricks}
\usepackage{epsfig}
\usepackage{pst-grad} % For gradients
\usepackage{pst-plot} % For axes
\usepackage{newlfont}
\usepackage{ifthen}
\usepackage{amsfonts}
\usepackage{amsthm}
\usepackage{graphicx}
\usepackage[breaklinks]{hyperref}
\usepackage{tikz}
\usepackage{subfigure}
\usepackage{float}

\definecolor{lime}{HTML}{A6CE39}
\DeclareRobustCommand{\orcidicon}{%
	\begin{tikzpicture}
	\draw[lime, fill=lime] (0,0) 
	circle [radius=0.16] 
	node[white] {{\fontfamily{qag}\selectfont \tiny ID}};
	\draw[white, fill=white] (-0.0625,0.095) 
	circle [radius=0.007];
	\end{tikzpicture}
	\hspace{-2mm}
}

\foreach \x in {A, ..., Z}{%
	\expandafter\xdef\csname orcid\x\endcsname{\noexpand\href{https://orcid.org/\csname orcidauthor\x\endcsname}{\noexpand\orcidicon}}
}

\begin{document}
\title{Model dielectric functions for fluctuation potential calculations in electron gas: a critical assessment}

\affiliation{Department of Physics, Birla Institute of Technology and Science-Pilani, Pilani Campus, Pilani,  Rajasthan, 333031, India}
\affiliation{Univ. Rennes, CNRS, IPR (Institut de Physique de Rennes) - UMR 6251, F-35000 Rennes, France}

\author{Aditi Mandal$^{1,2}$ \orcidA{}, Sylvain Tricot$^{2}$ \orcidD{},  Rakesh Choubisa$^{1}$ \orcidC{} and Didier S\'{e}billeau$^{2}$ \orcidB{}}
\email{didier.sebilleau@univ-rennes1.fr}


% \date{\today}

\begin{abstract}
In this article, we report a critical assessment of dielectric function calculations in electron gas through the comparison of different modelling methods. This 
work is motivated by the fact that the dielectric function is a key quantity in the multiple scattering description of plasmon features in various electron-based spectroscopies. 
Starting from the standard random phase approximation (RPA) expression, we move on to 
correlation-augmented RPA, then damped RPA models. Finally, 
we study the reconstruction of the dielectric function from its moments, using the Nevanlinna and memory function approaches. We find the memory function method to be the 
most effective, being highly flexible and customizable.

\end{abstract}

\maketitle 

%%%%%%%%%%%%%%%%%%%%%%%%%%%%%%%%%%%%%%%%%%%%%%%%%%%%%%%%%%%%%%%%%%%%%%%%%%%%%% 

\section{Introduction}

Plasmon effects occur in many electron-driven spectroscopies and are the signature of the response of the system to the sudden appearance of an extra charge (a traveling electron, a hole left behind, etc).
In core-level photoemission for instance,  plasmons appear as separate peaks at energies below the core peak. The energy differences with the core peaks are multiple of the plasmon energy $n\hbar\omega_{p}$. Here, $n$ is the number of plasmon losses suffered by the photoelectron before escaping the material under study. Values of $n$ at least up to
6 have been observed by Barman and coworkers~\cite{a17} in Aluminium. More recent results using 6 keV Hard X-ray photoelectron spectroscopy (HAXPES) seem to exhibit up to 14 plasmon peaks \cite{a17b}. For surface-sensitive photoemission,
the surface plasmon peak can be distinguished from the bulk one as they appear at different energies. 

The information embedded into plasmon peaks has not been much studied so far in photoemission. Back in 1990, Osterwalder and coworkers showed that they exhibited photoelectron diffraction-like features just as their core-level peak ~\cite{a54}. More recently, David, Godet and coworkers proposed to use these plasmon peaks to extract from their energy distribution some information on the system's dielectric function  ~\cite{a55, a56b}. They termed this new spectroscopy Photoemission Electron Energy Loss Spectroscopy (PEELS). Similarly, plasmon structures originating from valence band electrons have also been studied by Guzzo and coworkers ~\cite{a57, a58}. Therefore, although not used that much yet as system's information provider, plasmon structures seem a promising tool to extract some new information from spectroscopies, and in our case, from photoemission.

From the theoretical point of view, Lars Hedin and coworkers introduced two approaches in order to model
plasmon features in photoemission and X-ray absorption: the GW + cumulant expansion and the quasiboson model Hamiltonian method ~\cite{a57, a58, a61}. The two methods have been shown to be formerly equivalent by Vigil-Fowler et al ~\cite{a61}. The GW + cumulant expansion has been favoured by Reining and coworkers  ~\cite{a57, a58} to model plasmon structures originating from valence bands. They showed in particular that neither the DFT nor the GW were able to model the plasmon. On the other side, Takashi Fujikawa and co-workers have used Hedin's quasiboson model in order to incorporate it into the multiple scattering description of spectroscopies  ~\cite{a36,a37,a38}. Following Fujikawa's approach, it can be shown that, using reasonable approximations, the first plasmon peak cross-section can be written as the product of the core peak cross section times a loss function, which is very convenient as standard multiple scattering codes such as MsSpec ~\cite{a22} compute the core peak cross-section. Therefore, the implementation of the plasmon peak into a core-level code only involves the computation of this loss function. Fujikawa and coworkers have shown that this function can be expressed as \cite{a38}

\begin{equation}\label{eq1}
\int f_{c}(\mathbf{r})V^{i}(\mathbf{r})d\mathbf{r}    
\end{equation}

where $f_{c}(\mathbf{r})$ is a function involving the core hole wave function and the escaping photoelectron wave function. $V^{i}(\mathbf{r})$ is the so-called fluctuation potential that describes the excitation of a plasmon (by either the core
hole or the photoelectron). The only unknown in this approach is the fluctuation potential. Both Hedin and Fujikawa have used for $V^{i}(\mathbf{r})$ the analytical expressions derived either by Inglesfield ~\cite{a7} or by Bechstedt ~\cite{a11, a21} .
These descriptions are based on a 3D metallic-type material with a well identified surface, and for photoelectrons of low kinetic energies, originating from the vicinity of the surface. Therefore, they completely exclude lower dimensional systems (quantum dots, quantum wires, quantum wells, etc) or semiconductors, heterostructures and Dirac systems such as graphene and related layers or bilayers. Moreover, they are not suited to HAXPES experiments where the surface is basically overlooked by the escaping electron. 

In order to treat more diverse type of materials and spectroscopies involving more energetic electrons, we propose an alternative method to the problem of the description of the fluctuation potential. For this, we go back to the definition that was given by Hedin and coworkers. In reference ~\cite{a11} they define this potential as

\begin{equation}\label{eq2}
V^{\mathbf{q}}(\mathbf{r})=\left|\frac{V_{c}(\mathbf{q})}{\displaystyle{\frac{\partial\varepsilon(\mathbf{q},\omega)}{\partial\omega}\vert_{\omega=\omega_{\mathbf{q}}}}}\right|^{1/2}e^{i\mathbf{q}.\mathbf{r}}
\end{equation}

where $V_{c}(\mathbf{q})$ is the Fourier transform of the Coulomb potential and $\varepsilon(\mathbf{q},\omega)$ is the dielectric function of the system. The derivation is taken along the plasmon dispersion $\omega_{\mathbf{q}}$ which can be obtained by solving $Re[\varepsilon(\mathbf{q},\omega)]=0$. 

Through the use of the dielectric function, this definition offers us a convenient, flexible and general way to compute the fluctuation potential needed to evaluate the cross-section of plasmon peaks. Moreover,
being based on a proper description of the dielectric function, it makes a direct connection with the PEELS method developed by Godet, David and coworkers ~\cite{a55, a56b}, thereby reinforcing their idea that this spectroscopy could be used  ultimately to map somehow the bulk or surface dielectric functions.

It is not the purpose of this article to implement the calculation of the plasmon peak. This will be done in a forthcoming study. Rather here, we dedicate our work to reviewing and documenting different models of dielectric functions in order to assess their sensitivity to various parameters. In this view, we remark
that some spectroscopies such as photoemission do not resolve the momentum of the plasmon and therefore will need to integrate over it, while others such as Electron Energy Loss Spectroscopy (EELS) are plasmon momentum-sensitive, at least for small values of $\mathbf{q}$.

Ideally, we could obtain the dielectric function from an ab-initio electronic structure calculation. Unfortunately, most standard ab-initio codes only compute the dielectric function at $\mathbf{q} = 0$, as they essentially use it for optical spectroscopies. Non zero $\mathbf{q}$ values are generally not implemented, and when they are, the calculation uses a lot of computing resources. In order to bypass this problem, we have developed a computer code called MsSpec-DFM (Dielectric function module), which will be published as a separate module of MsSpec ~\cite{a22}. This new code 
can compute different model dielectric functions for a very large range of materials. The purpose of this article is to compare different approaches to analytically model a dielectric function and try to ascertain the most accurate. All the methods we use are based either on the homogeneous gas approach or on a Fermi liquid one. This means that the electron system that responds to the sudden appearance of either a core hole or a photoelectron has no structure whatsoever. Later, in a forthcoming work,
we will try to assess the effect of the band structure (non homogeneous distribution of $\mathbf{q}$) and of the crystal structure (interaction with phonons) on the plasmon description. Work is currently in progress using the Questaal LMTO code ~\cite{a63} to model a band structure-sensitive plasmon dispersion ~\cite{a64}, and preliminary results are very promising.

In section II, we recall the basics of the Hedin-Fujikawa quasiboson multiple scattering description
of the plasmon photoemission peak. Subsection A introduces the fluctuation potential and subsection B
the essentials on dielectric function formulation. Section III is devoted to the description of various model
dielectric functions encompassing: (i) the simple plasmon pole approximation, (ii) the RPA approximation,
(iii) the correlation-augmented RPA approximations. As these models do not conserve the number of particles,
we consider in section IV two approximations that ensure this conservation: (i) The Mermin approximation
and (ii) the Hu-O'Connell approximation. More refined methods that also conserve the momentum and the energy
such as the Atwal-Ashcroft \cite{a65} are not tested here because they can be easily implemented into
the family of dielectric functions described in the section V. Section V deals with a more unusual
pathway to build up a dielectric function. Indeed, as we mentioned before, most standard model dielectric
functions do not conserve the number of particles. In addition, the RPA does not contain correlation effects
and they have to be added externally through local field corrections (LFC).
It can be shown however that the $\omega^1$ moment of the inverse of the dielectric function is nothing else
than an expression of the conservation of the number of particles. Likewise the $\omega^3$ moment can be shown
to reflect the electron pair correlations \cite{a28} and the $\omega^5$ moment the electron triplet correlations \cite{a66}.
Consequently, if we can reconstruct the dielectric function, or its inverse, from its moments, we will
 will have "built-in" both the conservation of the number of particles and the correlations. We document here
 two methods based on this approach: (i) the Nevanlinna function method and (ii) the memory function
 method. We show that these two methods do improve upon the correlation-augmented RPA methods or the
 damping methods (Mermin, Hu-O'Connell). In addition, complex energies can be used that incorporate
 plasmon damping. Although in this article we will limit ourselves to the scalar memory function approach,
 we note that it can be augmented to a matrix version that will incorporate the conservation of the momentum
 and the conservation of the energy. As the memory function is based on the way a system relaxes, other features, such as the timescales of different plasmon decay modes,  can also be added into the memory matrix method which makes it the most flexible and general way to model "analytically" a dielectric function. Furthermore, it can be shown that most, if not all, other methods can be shown as particular cases. A comparative discussion is proposed in section VI.
 
 Note that all throughout this article, we consider the case of Aluminium for our calculations. This particular choice comes from the fact that Aluminium is a much studied system usually considered as a test case, or toy model.

\section{Theoretical background}

Following Hedin's quasiboson model Hamiltonian method ~\cite{a8,a11,a59} we add the coupling to
the plasmon field as a perturbation and treat it within
MS theory. The quasiboson model Hamiltonian in our case can be written as,
\begin{equation}
\begin{array}{ll}
H= \sum\limits_{q}\hbar\omega_{q}b_{q}^{\dagger}b_{q} \,+\, \sum\limits_{\mathbf{k}}\epsilon_{\mathbf{k}}c_{\mathbf{k}}^{\dagger}c_{\mathbf{k}} \, \\+ \, \sum\limits_{q\mathbf{k}\mathbf{k^{\prime}}}\Bigg[V^{q}_{\mathbf{k}\mathbf{k^{\prime}}}b_{q}^{\dagger} + (V^{q}_{\mathbf{k}\mathbf{k^{\prime}}})^{*}b_{q}\Bigg]c_{\mathbf{k}}^{\dagger}c_{\mathbf{k^{\prime}}} - \sum\limits_{q}V^{q}_{cc}(b_{q}+b_{q}^{\dagger})
\end{array}    
\end{equation}
where the first term describes the free boson field, the second term corresponds to the ejected  photoelectron, the third term corresponds to the interaction between the photoelectron and the boson field. The last term represents the core hole-boson coupling. In this approach, the interaction between an external charge and the electrons in the system is described through the fluctuation potential $V^{\mathbf{q}}$ given by Eq.~\ref{eq2} where $\omega_{\mathbf{q}} $ is the frequency of a plasmon of potential energy $\hbar\omega_{p}$ and momentum $\hbar\mathbf{q}$. The momentum vector of a plasmon is  given by $\mathbf{q}=\mathbf{k}_{in}-\mathbf{k}_{sc} $ where $ \mathbf{k}_{in}$ is the momentum of the electron before the plasmon loss and $\mathbf{k}_{sc}$ is the the momentum of the electron after the plasmon loss. 
 
 In their respective works, Hedin and coworkers and Fujikawa and coworkers relied on two analytical fluctuation potentials developed separately by Bechstedt~\cite{a21} and Inglesfield ~\cite{a20}, which we will describe later for the sake of completeness. 
 
 In the Hedin-Fujikawa formalism, the intensity of the first plasmon peak is given by ~\cite{a38},

\begin{equation}\label{eq5}
I^{1}(\mathbf{k},\hbar\omega_{\mathbf{q}},\hbar\omega_{i}) = I^{0}(\mathbf{k},\hbar\omega_{\mathbf{q}})\frac{\alpha(\hbar\omega_{i})}{\hbar\omega_{i}},
\end{equation}

where $\mathbf{k}$ is the momentum of the detected electron. The "no-loss" core-peak cross-section is ~\cite{a38}

\begin{eqnarray}\nonumber
I^{0}(\mathbf{k},\hbar\omega_{\mathbf{q}})= \, 4\pi^{2}\alpha_{FS} \, \frac{\hbar}{m^{2}\omega_{\mathbf{q}}}\mid<\tilde{\phi}_{\mathbf{k}}\mid\Delta\mid \, c>^{0}\mid^{2} \\
\, \times\exp\big[-\int_{0}^{+\infty}\frac{\alpha(\epsilon)}{\epsilon}d\epsilon \big]
\end{eqnarray}

The term before the multiplication sign is the 
usual core-level cross-section which is computed 
by standard MS codes such as MsSpec.

Within the Hedin-Fujikawa formalism, the loss function $\alpha(\epsilon) / \epsilon$ can be expressed as:

\begin{equation}\label{eq6}
\frac{\alpha(\epsilon)}{\epsilon} = \sum_i\mid\int f_{c}(\mathbf{r})V^{i}(\mathbf{r})d\mathbf{r}\mid^2 \delta(\epsilon-\hbar\omega_{i})
\end{equation} 

where $f_{c}(\mathbf{r})$ is a well-defined  core-state related function and $V^{i}(\mathbf{r})$ is the fluctuation potential corresponding to the excitation of a plasmon of energy $\hbar\omega_{i}$. So, for quantitative modelling of plasmon features in spectroscopies it is imperative to have a good fluctuation potential as it is the only unknown in \eqref{eq6}. Both Hedin and Fujikawa have used for $V^{i}(\mathbf{r})$ the analytical expressions derived either by Inglesfield ~\cite{a7} or by Bechstedt ~\cite{a11, a21}. These descriptions are based on a 3D metallic-type material with a well identified surface, and for photoelectrons of low kinetic energies, originating from the vicinity of the surface. Therefore, they completely exclude lower dimensional systems (quantum dots, quantum wires, quantum wells...) or semiconductors, heterostructures and Dirac systems such as graphene and related layers or bilayers. Moreover, they are not suited to HAXPES experiments where the surface is basically overlooked by the escaping electron. In order to treat more diverse type of materials and spectroscopies involving more energetic electrons, we propose an alternative method to the problem of the description of the fluctuation potential.

\subsection{Fluctuation potentials}
The fluctuation potential describes the coupling between the electron and the bosons. It is not the purpose of this article to go into the details of the analytical derivation of the ones that can be found in the 
literature. From equation \eqref{eq2}, we see that it can be factorized as

\begin{equation}\label{eq9a}
V^{\mathbf{q}}(\mathbf{r})=V^{\mathbf{q}} \, e^{i\mathbf{q}.\mathbf{r}}
\end{equation}

The different models of analytical fluctuation potentials available in the literature have been obtained within linear response theory. In addition, in the case of Inglesfield and Bechstedt, the choice of a semi-infinite electron gas implies that  the exponential in \eqref{eq9a} reduces essentially to the form
$\exp \left[ \mathbf{q}_\parallel \; z \right]$.

\subsubsection{Plasmon-pole}

The plasmon-pole fluctuation potential was  originally derived by Lundqvist [19]. Its value is

\begin{equation*}
\left\{
\begin{array}{lll}
V^{\mathbf{q}}(z)& = & V^{\mathbf{q}}_{\mathrm{PP}} \; e^{i\mathbf{q}.\mathbf{r}}\\
& & \\
V^{\mathbf{q}}_{\mathrm{PP}} & = & \sqrt{V_{c}(\mathbf{q})\dfrac{\omega_{p}^{2}}{\omega_{\mathbf{q}}}\dfrac{1}{2V}}
\end{array}
\right.
\end{equation*}

where $V^{\mathbf{q}}_{\mathrm{PP}} $ is the electron plasmon coupling constant and $V$ is the system volume. For the surface case, we replace the 3D Fourier transformation of the bare Coulomb potential $V_c(\mathbf{q})$ by the 2D transformation and take into account that $\mathbf{q}$ is parallel to the surface.
 
\subsubsection{Inglesfield}

Inglesfield introduced a more realistic analytical potential, using the fact that the bulk modes in a semi-infinite system are standing waves, i.e., phase-shifted cosines modified at the surface ~\cite{a20}. The Inglesfield bulk plasmon potential is given as:

\begin{equation}
\begin{array}{ll}
V^{\mathbf{q}}(z) = &
\displaystyle{V^{\mathbf{q}}_{\mathrm{IN}}[\cos(-q_{z}z+\phi_{\mathbf{q}})-e^{q_{\parallel}z}\cos\phi_{\mathbf{q}}]\theta(z)}, \, \text{inside}, \\
\\
\,\,\,\,\,\,\,\,\,\, \, \, \, \, \, \, \,  = & 0, \, \,\text{outside}. 
\end{array}
\end{equation} 

where

\begin{equation}
\begin{array}{lll}
\displaystyle{V^{\mathbf{q}}_{\mathrm{IN}}} & = & \displaystyle{\sqrt{\frac{\omega_{p}^{2}}{\omega_{\mathbf{q}}}\frac{1}{V}V_{c}(\mathbf{q})}}, \\
& & \\
\displaystyle{\phi_{\mathbf{q}}} & = & \displaystyle{\tan^{-1}(\frac{q_{\parallel}}{q_{z}})}
\end{array}    
\end{equation}

Similarly, Inglesfield surface plasmon potential is given as:

\begin{equation}
\begin{array}{lll}
\displaystyle{V^{\mathbf{q}}}(z) & = & \displaystyle{V^{\mathbf{q}}_{\mathrm{IN}} \; e^{q_{\parallel}z}}, \\
& & \\
\displaystyle{V^{\mathbf{q}}_{\mathrm{IN}}} & = & \displaystyle{\sqrt{V_{c}^{2D}\frac{\omega_{p}}{A\sqrt{8}}}}
\end{array}    
\end{equation}
The Coulomb potential Fourier transforms being
\begin{equation}\label{eq10}
V_{c}= \, V_{c}^{2D} \, = \dfrac{e^{2}}{2\epsilon_{0}\sqrt{q^{2}+k_{s}^2}}
\end{equation}

where $k_s$ is a screening momentum.

\subsubsection{Bechstedt}

Bechstedt and co-workers  derived an expression for the screened potential $W$. It was recast in terms of the fluctuation potential by Hedin and coworkers ~\cite{a11}:
\begin{equation}
\begin{array}{lll}
V^{\mathbf{q}}(z) =& \displaystyle{N_{b} \, [ e^{q_{\parallel}z}\theta(z) + \lbrace(2+C_{1}+C_{3})\, \cos(q_{z}z)}\\
\\
& \displaystyle{-C_{2}\, \sin(q_{z}z)-(1+C_{1})e^{q_{\parallel}z}-} \\
\\
& \displaystyle{C_{3}\exp(\sqrt{\omega_{p}+\omega_{\mathbf{q}}+q_{\parallel}^{2}}z)\rbrace \, \theta(-z)]} 
\end{array}    
\end{equation}

where the coefficients are given by
\begin{equation}
\begin{array}{lll}
C_{1}= \dfrac{\omega_{p}^{2}}{\omega_{\mathbf{q}}^{2}-\omega_{p}^{2}}, \\
C_{2}= \dfrac{q_{\parallel}\omega_{p}^{2}}{2\omega_{\mathbf{q}}(\omega_{\mathbf{q}}-\omega_{p})q_{z}}, \\
C_{3}= \dfrac{q_{\parallel}\omega_{p}^{2}}{2\omega_{\mathbf{q}}(\omega_{\mathbf{q}}-\omega_{p})q_{z}\sqrt{\omega_{p}+\omega_{\mathbf{q}}+q_{\parallel}^{2}}}
\end{array}    
\end{equation}

In the surface case,

\begin{equation}\label{eq11}
V^{\mathbf{q}}(z)= N_{s}\left[e^{-q_{\parallel}z}\theta(z)+a(q_{\parallel},z,\omega)\theta(-z)\right] 
\end{equation}

where $a(q_{\parallel},z,\omega) $ is related to the bulk dielectric function through ~\cite{a21}

\begin{equation}\label{eq12}
a(q_{\parallel},z,\omega)=\frac{2q_{\parallel}}{\pi}\int_{0}^{\infty}\frac{\cos(q_{z}z)}{q^{2}\varepsilon(\mathbf{q},\omega)}dq_{z}
\end{equation}

where $\varepsilon(\mathbf{q},\omega)$ can be computed within the simple plasmon-pole model.

\begin{figure}[htp]
\includegraphics[width = 0.9\columnwidth]{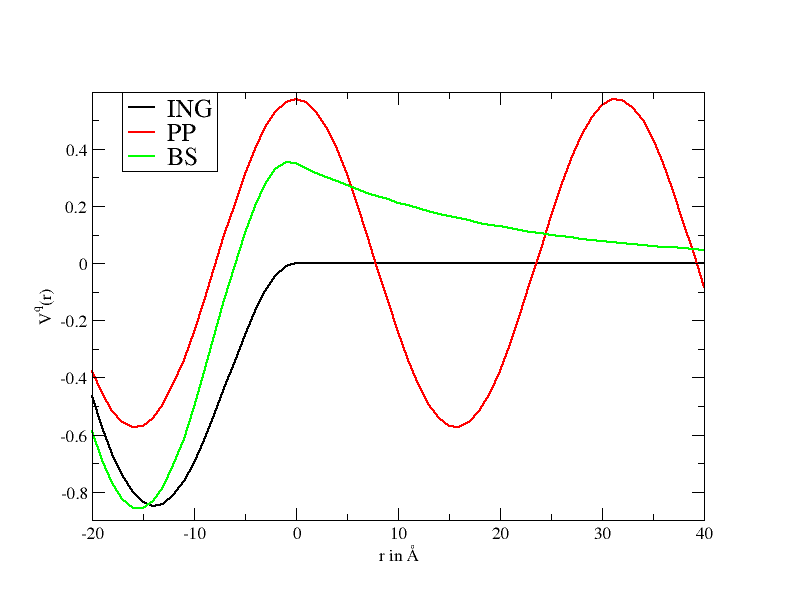}
\caption{$r$-variations of the standard analytical fluctuation potentials.}\label{fig1}
\end{figure}

These three potentials are the standard analytical fluctuation potentials available in literature. Their space variations are represented in Fig.\ref{fig1}, in the Aluminium case. However, due to the fact that they are surface-related (and derived for metals), they cannot be used to describe spectra derived for reduced-symmetry
systems such as quantum dots, quantum wells, graphene and other 2D materials, Dirac materials, etc. 
In addition, they are not suited to high-energy 
spectroscopies such as HAXPES where the surface can be safely ignored.

In order to overcome the limitations of these 
potentials, we will go back to Hedin's definition  of the fluctuation potential (equation \eqref{eq2}), which expresses this potential 
as a function of the dielectric function. 
This will allow us a much more flexible approach where the true dimensionality and structure of the material can be properly taken into account.

\subsection{Dielectric function: General Background}
 
The dielectric function of a system describes the response of this system to an external perturbation. It is determined by the properties of the system and its interaction with the perturbing object. Several related quantities can be found in the literature, together with their relationship with various spectroscopies. 
for instance, it is well-known that the cross-section of the EELS is related to the loss function. Therefore, in the following subsection, we will introduce the different 
quantities of interest for spectroscopies and that are related to the dielectric function. 

We start by decomposing the dielectric function 
into its real and imaginary parts

\begin{equation}\label{eq13}
\varepsilon(\mathbf{q},\omega)= \varepsilon_{1}(\mathbf{q},\omega)+ i\varepsilon_{2}(\mathbf{q},\omega)
\end{equation}

Then, the loss function $L(\mathbf{q}, \omega)$ is related to dielectric function of the solid through,

\begin{equation}\label{eq14}
L(\mathbf{q}, \omega)\, \propto\, \Im \left[ \varepsilon^{-1} (\mathbf{q}, \omega)\right]
\end{equation}

Similarly, the dynamical structure factor 
can be expressed as ~\cite{a50}

\begin{equation}\label{eq15}
S(\mathbf{q}, \omega) = \Im \left[\frac{-1}{\varepsilon(\mathbf{q}, \omega)}\right] = \, \frac{\varepsilon_{2}(\mathbf{q}, \omega)}{\mid\varepsilon(\mathbf{q}, \omega)\mid^{2}}
\end{equation}

It describes the spectrum of excitations in the systems as a function of the momentum transfer 
$\mathbf{q}$ and the energy transfer.

Likewise, the susceptibility, or density-density response function,  can be defined as

\begin{equation}\label{eq16}
\chi(\mathbf{q}, \omega) = \dfrac{1}{V_{c}(q)}\left[\dfrac{1}{\varepsilon(\mathbf{q}, \omega)}-1 \right]
\end{equation}

With these tools, we can access many different ways to model the dielectric function,  and compute cross-sections.

\subsubsection{The simplest case: Plasmon pole}

The plasmon pole dielectric function describes the response of the system entirely in terms of collective modes. The dielectric function is just the analytic continuation of a simple pole and is given by ~\cite{a39}

\begin{equation}\label{eq17}
\varepsilon(\mathbf{q},\omega)= 1 - \frac{\omega_{p}^{2}}{\omega^{2}+\omega_{p}^{2}-\omega_{q}^{2}}
\end{equation}

Its real part and imaginary part are represented in 
figure \ref{fig:foobar} for the case of Aluminium.

The plasmon dispersion band corresponds to the upper band of  $\Re[\varepsilon(\mathbf{q},\omega)]=0$.

\section{RPA and beyond}

\begin{figure*}
    \centering
    \subfigure{\includegraphics[width=0.32\textwidth]{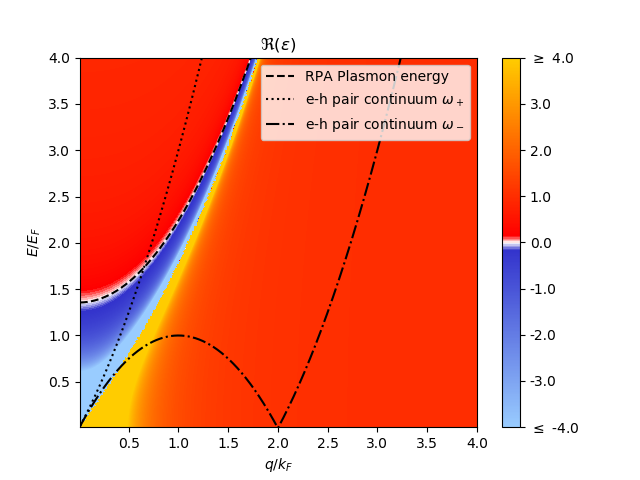}} 
    \subfigure{\includegraphics[width=0.32\textwidth]{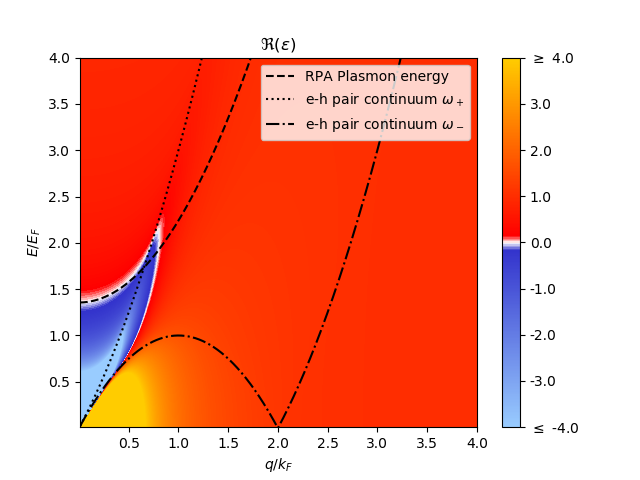}} 
    \subfigure{\includegraphics[width=0.32\textwidth]{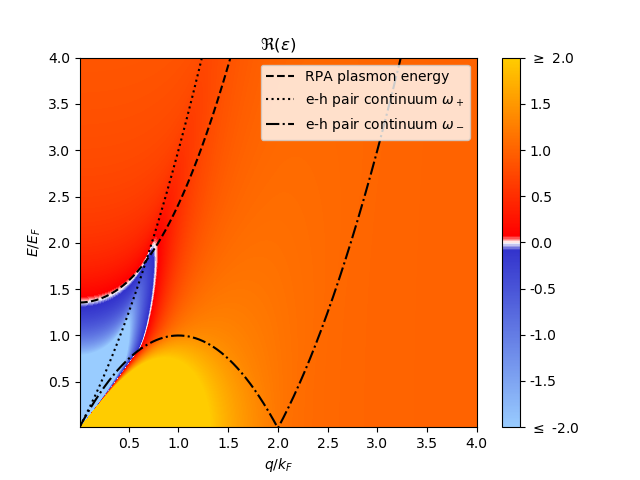}}
    \subfigure{\includegraphics[width=0.32\textwidth]{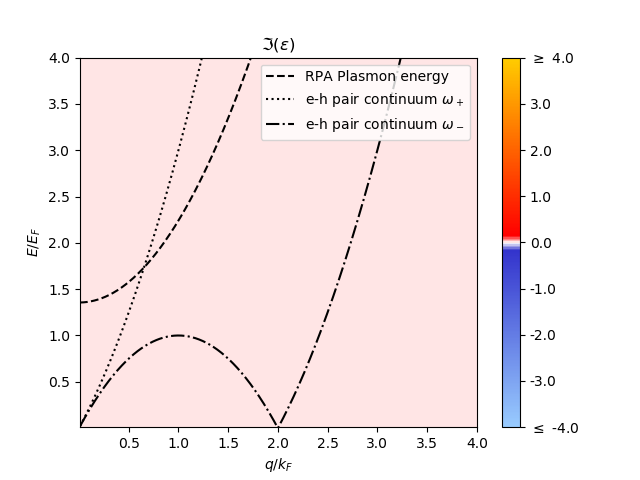}}
    \subfigure{\includegraphics[width=0.32\textwidth]{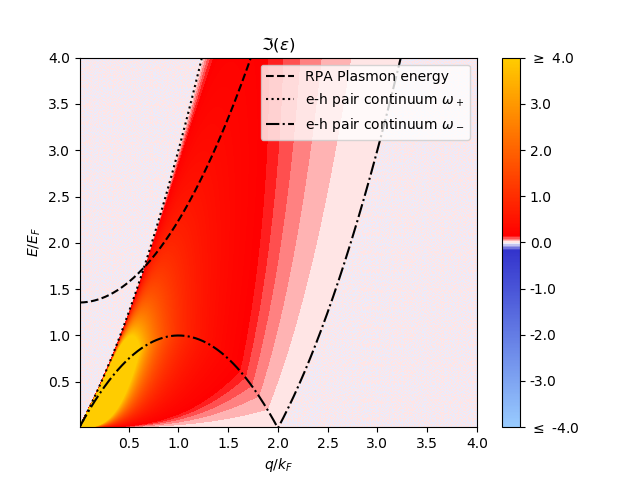}}
    \subfigure{\includegraphics[width=0.32\textwidth]{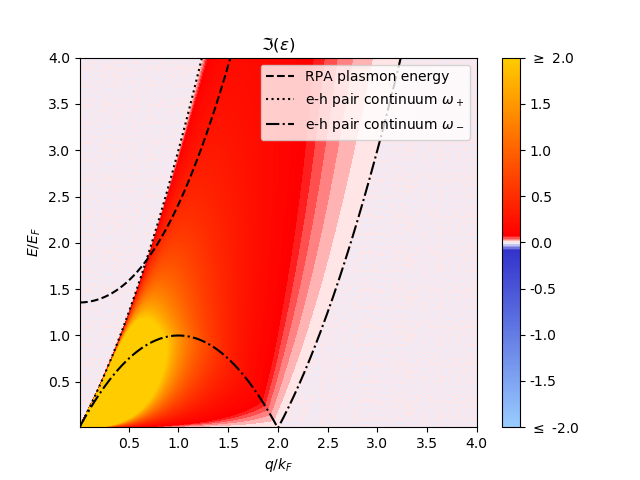}}
    \caption{Real (above) and imaginary (below) part of dielectric function $\epsilon(\mathbf{q}, \omega)$ for (left) plasmon pole, (center) RPA,  (right) RPA + UTI1  respectively. }
    \label{fig:foobar}
\end{figure*}

The plasmon pole approximation's main drawback, as can be seen from the left-hand part of figure \ref{fig:foobar}, is that it does not incorporate any damping mechanism. Therefore, we have a plasmon (the upper white band in the real part of $\varepsilon(\mathbf{q},\omega)$) which, once created, never decays. This is clearly not physical. Historically, the first expression of the dielectric function to include a damping mechanism is the 
Random Phase approximation (RPA) originally derived by Lindhard~\cite{a23}. This approximation is based on the description of the delocalized electrons system as a \emph{homogeneous and non-interacting electron gas}. The real and imaginary parts 
of the RPA dielectric function are represented in the  middle figure of figure \ref{fig:foobar}. Now, we have a clear damping inside the so-called Landau region which is delimited by the two parabola $\omega_+$ and $\omega_-$. This damping comes from the decay of the plasmon into an electron-hole pair. There is however no damping mechanism built-in outside the Landau region, which means that there, the plasmon will "live" forever. 
In addition, being built on the assumption that the electron gas is  non-interacting, the RPA does not allow for correlation effects.  

Correlation effects can be introduced into the RPA through the so-called \emph{local field corrections} (LFC) $G(\mathbf{q},\omega)$. These corrections are related to the exchange and correlation term of the Density Functional Theory (DFT). For a given correction term, the correlation-augmented dielectric function can be written as

\begin{equation}\label{eq18}
\varepsilon(\mathbf{q},\omega)=\frac{V_{c}(\mathbf{q})\Pi^{RPA}(\mathbf{q},\omega)}{1+V_{c}(\mathbf{q})G(\mathbf{q},\omega)\Pi^{RPA}(\mathbf{q},\omega)}
\end{equation}

where $G(\mathbf{q},\omega)$ is the dynamical local field correction. The RPA dielectric function is given by

\begin{equation}\label{eq19}
\varepsilon^{RPA}(\mathbf{q},\omega)=1-V_{c}(\mathbf{q})\Pi^{RPA}(\mathbf{q},\omega)
\end{equation}

where $\Pi^{RPA}(\mathbf{q},\omega)$ is the RPA polarization.

Very few models of \emph{dynamical} local field corrections exist in the literature, while a lot of attention has been 
devoted to \emph{static} corrections $G(\mathbf{q})$. Therefore, we will restrict here to the latter.

In order to investigate the effect of such corrections on the behaviour of the dielectric function, we consider here 
three different types of static corrections, the Hubbard model (HUBB), which takes only into account exchange effect, the 
Pathak-Vashista correction (PVHF) \cite{a28}, and the 
Utsumi-Ichimaru (UTI1) one \cite{a30}. They can be expressed respectively as 

\begin{equation}\label{eq20}
\left\{
\begin{array}{lll}
G(\mathbf{q})_\mathrm{HUBB} & = & \displaystyle{ \frac{1}{2}\frac{x^{2}}{1 + x^{2}}} \\
& & \\
G(\mathbf{q})_\mathrm{PVHF} & = & \displaystyle{\frac{1}{\omega_{p}^2} \; J(q)} \\
& & \\
G(\mathbf{q})_\mathrm{UTI1} & = & \displaystyle{\frac{3(4-x^2) \, (28+5x^{2})}{16x}}\times \log\left\vert\dfrac{2+q}{2-q}\right\vert\Bigg]
\end{array}
\right.
\end{equation}

where we have used the notation $x = q / k_F$.

Here, the Kugler function $J(q)$ is given by \cite{a29}

\begin{equation}\label{eq22}
J(q) = \, \frac{e^2}{m \pi} \; \int_0^{+ \infty} \, k^2 \; \left[ S(k) -1 \right] \; J(k,q) \; dk
\end{equation}

where $J(k,q)$ is given by

\begin{equation}\label{eq23}
J(k,q) = \, \frac{5}{6} - \frac{k^2}{2 q^2} + \frac{q}{4k} \left( \frac{k^2}{q^2} - 1 \right)^ 2 
\; \ln \left| \frac{k + q}{k - q } \right|
\end{equation}

$S(q)$ is the static structure factor. Here, we  approximate it by its Hartree-Fock value

\begin{equation}\label{eq25}
\begin{array}{llll}
S_{HF}(q) & = & \displaystyle{\frac{3}{4} \; \frac{q}{k_F} - \frac{1}{16} \; \left( \frac{q}{k_F} \right)^3} &  \text{for } q < 2 k_F \\
 & & & \\
  & = & 1 & \text{for }  q > 2 k_F 
\end{array}
\end{equation}

\begin{figure}[htp]
\includegraphics[width = 0.9\columnwidth]{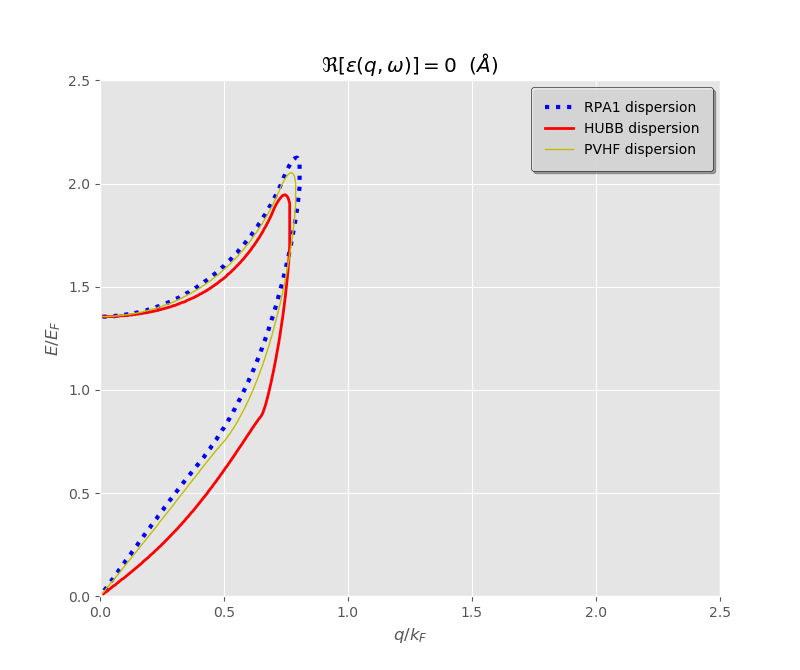}
\caption{Comparison of collective excitations dispersion lines for  RPA and some RPA + local field corrections. The plasmon dispersion corresponds to the upper band.}\label{fig3}
\end{figure}

The right-hand plots of Fig.\ref{fig:foobar} corresponds to the real part (up) and imaginary part (down) of the 
UTI1-enhanced RPA dielectric function. We see clearly a change in the dispersion of the collective excitations (white band), with respect to the RPA case. The comparison 
between this dispersion is even clearer in figure \ref{fig3} where we give a 2D representation of the collective excitations bands for the RPA, RPA + Hubbard and RPA + Pathak-Vashista static local field corrections. Nevertheless, even if we now incorporate 
correlation effects into the dielectric function, we see clearly that, despite some minor changes in the imaginary 
part of the dielectric function (see Fig.\ref{fig:foobar}), 
we are still unable to properly describe plasmon damping outside the Landau regime. In principle, we could go to 
dynamic LFC, but in the following, we will rather explore 
completely other ways to describe the dielectric function 
with built-in damping.

\section{Damping-based dielectric functions}

RPA alone has a certain number of limitations:  (i) it fails to conserve the number of particles, (ii) it does not contain correlation effects (they have to be added externally through local field corrections) and (iii) it does not incorporate plasmon damping outside the Landau regime. In order to cure (i), Mermin \cite{a40} extended the Lindhard dielectric function in the relaxation-time approximation, where  essentially the collisions relax the electronic density matrix not to its uniform equilibrium value, but to a local equilibrium density matrix. The Mermin ~\cite{a41} dielectric function thus can be written as:

\begin{equation}\label{eq27}
\varepsilon(\mathbf{q},\omega) \, = \, 1 + \frac{\displaystyle{\left( 1 + \frac{i}{\omega \tau} \right) }\;
\left[ \varepsilon^0(\mathbf{q},\omega + i/\tau) - 1 \right]}{1 + \displaystyle{\frac{i}{\omega \tau} \;
\left[ \frac{\varepsilon^0(\mathbf{q},\omega + i/ \tau) - 1}{\varepsilon^0(\mathbf{q},0) - 1} \right]}}
\end{equation}

where $\varepsilon^0(\mathbf{q},\omega)$ is the RPA dielectric function and $\bar{\omega} = \omega + i/\tau$ is a complex frequency incorporating damping through the relaxation time $\tau$. Mermin's interest here is in the long wavelength limit behaviour of the electron gas where the focus is on obtaining a modified Lindhard dielectric function which reduces to the correct classical behaviour in the $q\longrightarrow0$ limit. 

\begin{figure*}[htp]
    \centering
	\subfigure{\includegraphics[width=0.9\columnwidth]{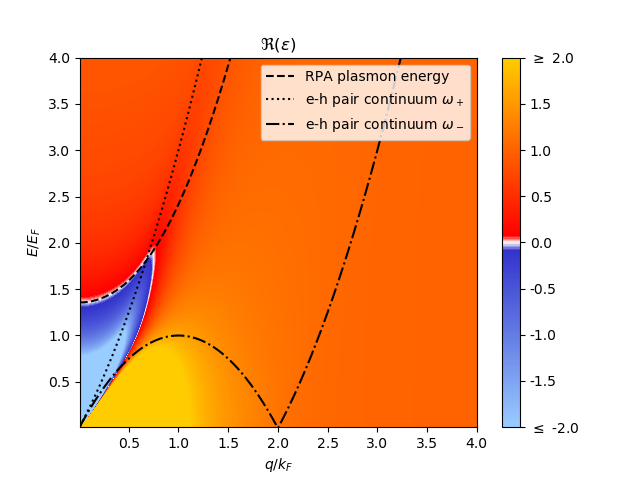}}
    \subfigure{\includegraphics[width=0.9\columnwidth]{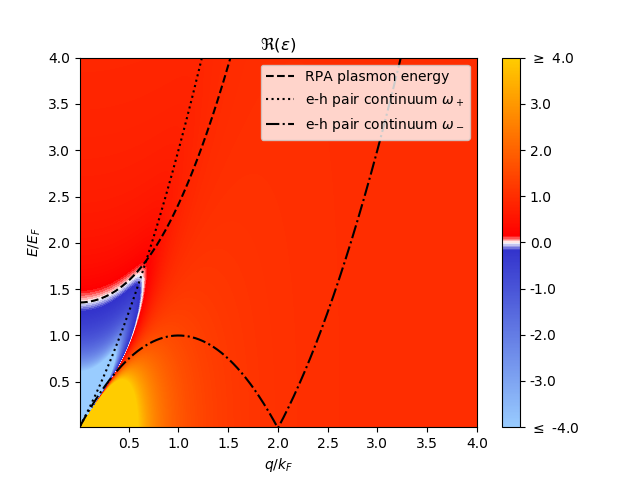}} 
    \subfigure{\includegraphics[width=0.9\columnwidth]{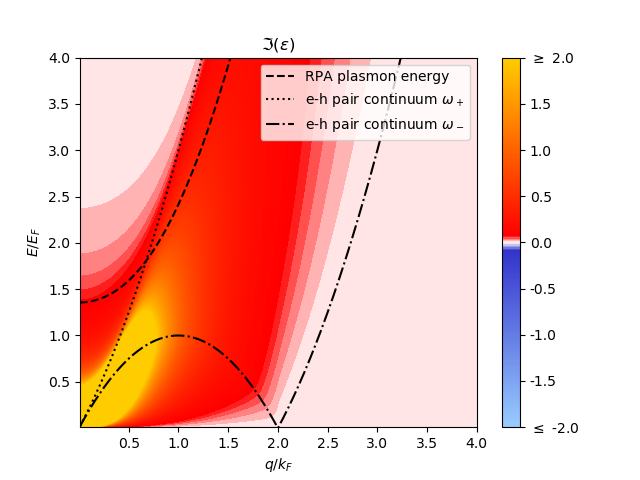}} 
    \subfigure{\includegraphics[width=0.9\columnwidth]{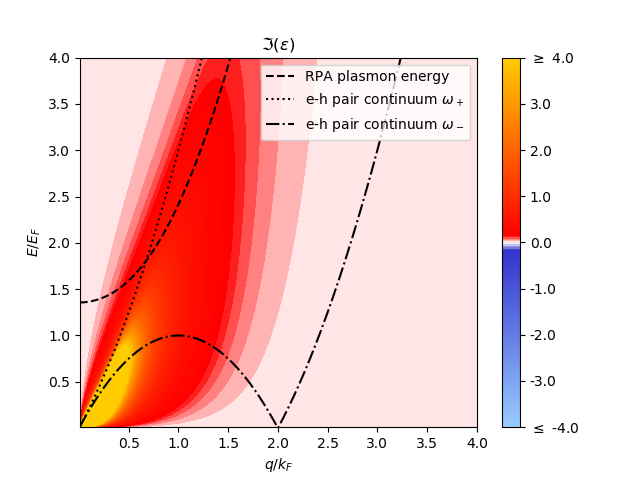}}
    \caption{Real and imaginary part of the (left) Mermin dielectric function with a relaxation time of $0.5$ fs, (right) Hu-O'Connell dielectric function with a diffusion coefficient $D = 2.9 10^{-5}$ m$^2$ s$^{-1}$ }\label{fig4}
\end{figure*}

Following the same idea, Hu and O' Connell ~\cite{a42} generalized the Lindhard dielectric function to include fluctuation effects arising from electron-electron and electron-impurity interactions. They also studied Friedel oscillations as an example of the application of their generalized version of the Lindhard function and observed that these oscillations are damped due to the inclusion of fluctuation effects. Due to the complexity of the Hu-O'Connell equations, which are expressed in terms of the diffusion coefficient $D$, we do not reproduce them here but refer instead to their article ~\cite{a42}. As displayed in Fig.~\ref{fig4}, we now observe a damping in the imaginary part of both the Mermin and the Hu-O'Connell dielectric function in the non Landau region. This damping can be regarded as the signature of a built-in lifetime for the plasmon. However, being based upon the standard RPA model, they lack a proper description of correlations. Refer to \ref{fig4} for the real and the imaginary part of both the Mermin and the Hu-O'Connell dielectric function.

These two approaches are encouraging, as we do need a proper damping of the plasmon in order to correctly describe the relaxation of the system. This damping is clearly absent in the standard plasmon pole and RPA methods, even when the latter is augmented with various \emph{static} local field corrections in order to incorporate electron-electron correlations.

\section{Dielectric function: Alternative approach}

An alternative family of dielectric functions can be obtained through a reconstruction from the first few moments. Two independent approaches can be found in the literature, the Nevanlinna approach ~\cite{a43} and the memory function approach ~\cite{a44}. The main advantage of these two approaches is that conservation of the number of particles and correlation are built-in.
We note that for larger values of $\omega$, $\Im \left[ \varepsilon^{-1} (\mathbf{q}, \omega)\right]$ tends quickly to zero, so that we have the expansion 
\begin{equation}\label{eq28}
\varepsilon^{-1} (\mathbf{q}, \omega) \, = \, 1 + \sum_{n = 1}^{+ \infty} \,
\frac{\left\langle {\omega^{2n-1}}\right\rangle_L}{\omega^{2n}}
\end{equation}
As mentioned in the introduction, it is well-known that the RPA dielectric function does not satisfy the compressibility sum rule and the frequency moment sum rules ~\cite{a46}. This realization is the starting point of the two above-mentioned approaches which we outline now.

\subsection{The Nevanlinna function method} 

The Nevanlinna approach is  based on the moments of the loss function. The standard loss function described in Eq.~\ref{eq14} is related to \emph{Nevanlinna loss function}% through,
\begin{equation}\label{eq29}
\mathcal{L}(\mathbf{q}, \omega) \, = \, \frac{L(\mathbf{q}, \omega)}{\omega}
\end{equation}
The corresponding moments are defined as ~\cite{a43}
\begin{equation}\label{eq30}
C_n \; = \, \frac{1}{\pi} \;  \int_{- \infty}^{+ \infty} \, \omega^{n-1} \;
\Im \left[ \frac{-1}{\varepsilon(\mathbf{q}, \omega)} \right] \; d \omega \, = \, \frac{1}{\pi} \; \left\langle {\omega^{n-1}}\right\rangle _\mathcal{L}
\end{equation}

A dimensional analysis shows that moments $C_2(q)$ has the dimension of a squared frequency, and likewise for every ratio $C_{n+2}(q)/C_n(q)$. Therefore, we introduce the notation
\begin{equation}\label{eq31}
\omega_n^2(q) \, = \, \frac{C_{2n}(q)}{C_{2n-2}(q)}
\end{equation}
The $\omega_n(q)$ have the dimension of a frequency and are the characteristic frequencies in the Nevanlinna function approach. 
The same characteristic frequencies will appear in the memory function method. The connection between the different types of moments can be made using, for the $T = 0$ K case,
\begin{equation}\label{eq32}
\begin{array}{lcrl}
L\left(\mathbf{q}, \omega \right) \, = \, \Im \left[ \displaystyle{ \frac{- 1}{\varepsilon(\mathbf{q}, \omega)}}\right] \, & = \, V_C(q) \; \displaystyle{\frac{\pi \bar{n}}{\hbar}} \;
S(\mathbf{q}, \omega) \, \\
& & & \\
 &= \, - V_C(q) \; \Im \left[ \chi(\mathbf{q}, \omega) \right]
\end{array} 
\end{equation}
where we find
\begin{equation}\label{eq33}
 V_C(q) \; \frac{\pi \bar{n}}{\hbar} \, = \, \frac{\pi}{2} \; \frac{\omega_p^2}{\omega_{\mathbf{q}}}
\end{equation}
implying
\begin{equation}\label{eq34}
\left\langle {\omega^{n}}\right\rangle _S \, = \, \frac{2}{ \pi} \; \frac{\omega_{\mathbf{q}}}{\omega_p^2}  \; \left\langle{\omega^{n}}\right\rangle \, = \,
-  \frac{\hbar}{\bar{n}\pi} \; \left\langle\omega^{n}\right\rangle\chi
\end{equation}
with $\bar{n}$ being the number density. Now, the frequency moments of the loss function can be written as ~\cite{a45},

\begin{equation}\label{eq35}
\left\langle\omega^{2n-1}\right\rangle_L \, = \, 2 \;  \int_0^{+ \infty} \, \omega^{2n-1} \;
\Im \left[ \frac{- 1}{\varepsilon(\mathbf{q}, \omega)} \right] \; d \omega
\end{equation}

So, then we have

\begin{equation}\label{eq36}
\begin{array}{llll}
\left\langle\omega^{-1}\right\rangle_L  & = & \pi \; \left[ 1 - \displaystyle{\frac{1}{\epsilon(q)}} \right]  \text{ compressibility sum rule} \\
& & & \\
\left\langle\omega\right\rangle_L  & = & \pi \; \omega_p^2  \; \; \;f\text{-sum rule} \\
& & & \\
\left\langle\omega^{3}\right\rangle_L & = & \pi \;  \displaystyle{ \omega_p^2 \; \left[ \omega_{\mathbf{q}}^2
+ 4 \; \omega_{\mathbf{q}} \frac{\left\langle{t}\right\rangle}{\hbar} + \omega_p^2 +  J(q)  \right] } \\
& 
\end{array}
\end{equation}

$J(q)$ is the Kugler function defined in equation \eqref{eq22} and $\left\langle{t}\right\rangle$ is the average 
kinetic energy per electron. 

$n=1$ ($f$-sum rule) ensures the conservation of the number of particles and $n=3$ ensures a proper account of two-body correlation effects.
The structure factor $S(\mathbf{q}, \omega)$ for a non-zero temperature system can be obtained from  the loss function through the fluctuation-dissipation theorem
\begin{equation}\label{eq37}
\Im \left[ \frac{-1}{\varepsilon(\mathbf{q}, \omega)} \right] \, = \,
\frac{\pi}{\hbar} \; V(\mathbf{q}) \;
\left[ 1 - \exp \left( - \displaystyle{ \frac{\hbar \omega}{k_B T}} \right) \right] \;
S(\mathbf{q}, \omega)
\end{equation}
For low energies, or large temperatures, this can be approximated by,
\begin{equation}\label{eq38}
\Im \left[ \frac{-1}{\varepsilon(\mathbf{q}, \omega)} \right] \, \approx \,
\frac{\pi}{k_B T} \; V(\mathbf{q}) \; \omega \; S(\mathbf{q}, \omega)
\end{equation}

The Nevalinna method is mathematically involved. It relies on the mathematical
 solution of the so-called  \emph{non-canonical solution of the Hamburger moment problem}. It
 is not the purpose of the present article to go into the details of the method. We refer
 the readers interested by this approach to the review article by Tkachenko ~\cite{a67}. The
 main point is that it involves a reconstruction of the dielectric function in terms of
 characteristic frequencies $\omega_n$ that are functions of the first moments of the
 loss function. Building up on the mathematical theorems, one arrives at the equation

\begin{equation}\label{eq39}
\varepsilon^{-1} (\mathbf{q}, \bar{\omega}) \, = \, 1 +
\frac{\omega_p^2 \; \left( \bar{\omega} + Q_2(\mathbf{q}, \bar{\omega}) \right) }{
\bar{\omega} \; (\bar{\omega}^2 - \omega_2^2) + Q_2(\mathbf{q}, \bar{\omega}) \;
(\bar{\omega}^2 - \omega_{1}^2)} \,
\Im \left[ \bar{\omega} \right] \ge 0
\end{equation}

for the 3-moment expression. Here, $\omega_p$ is the plasmon frequency,
$Q_2(\mathbf{q} ,\omega)$ is the \emph{unknown} Nevalinna function and $\bar{\omega}$ is the complex
frequency whose imaginary part represents the plasmon damping. We note that
$Q_2(\mathbf{q}, \bar{\omega}) $ has the dimension of a frequency. 

\begin{figure}[htp]
\includegraphics[width = 0.9\columnwidth]{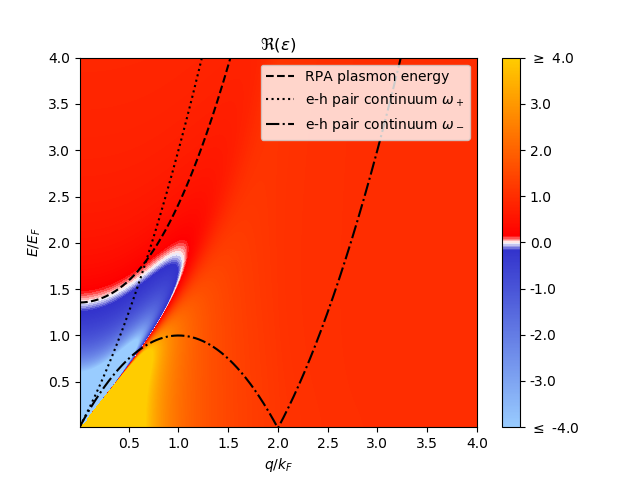}
\includegraphics[width = 0.9\columnwidth]{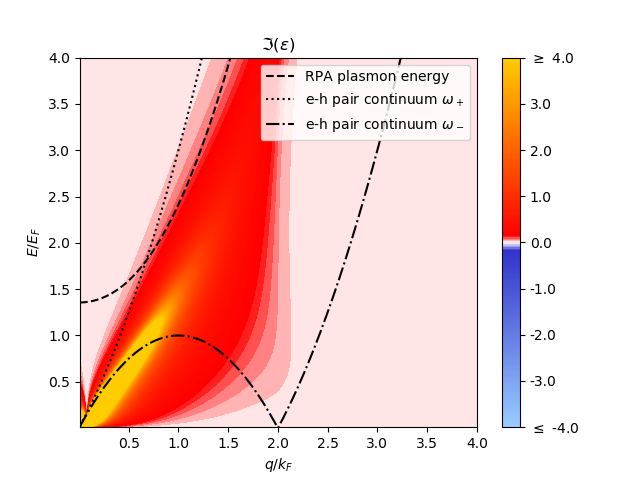}
\caption{Real and imaginary part of the 3-moment Nevanlinna dielectric function within the static approximation STA3.}\label{fig5}
\end{figure}

In an electron liquid, the Nevanlinna function plays the role of the \emph{dynamical} local field correction
$G(\mathbf{q} ,\omega)$. More precisely, we have ~\cite{a68}

\begin{equation}\label{gq}
G({\mathbf q}, \bar{\omega}) \, = \, 1 + \frac{1}{\left[ \varepsilon_\mathrm{RPA}(\mathbf{q}, \bar{\omega})  - 1 \right]}
+ \frac{\bar{\omega}^2}{\omega_p^2} -
\frac{\bar{\omega} \omega_2^2 + \omega_1^2 \; Q_2(\mathbf{q}, \bar{\omega})}{\omega_p^2
\left( \bar{\omega} + Q_2(\mathbf{q}, \bar{\omega}) \right)}
\end{equation}

Nevalinna functions must fulfill a number of mathematical properties including having a Riesz-Herglotz representation ~\cite{a47, a67}.
Therefore, finding relevant functions is a complex mathematical problem that has led to several papers in the literature \cite{a67,a68,a43b,a48}.

Here, we make the choice ~\cite{a48}

\begin{equation}\label{eq40}
Q(q,\bar{\omega}) \, = \, i \; \frac{\pi}{2} \; \varepsilon(q) \; \left(  \varepsilon(q) - 1 \right) \; 
(q a_0) \; \omega_{\mathbf{q}} \; 
\left[ \frac{\omega_2^2}{\omega_1^2} - 1 \right]
\end{equation}

which is well suited to our needs.

From Fig.\ref{fig5}, we observe that the correlations are properly taken into account (there is a dramatic change into the plasmon dispersion with respect to RPA-based methods) and in the imaginary part of the Nevanlinna dielectric function we see some damping of the plasmon outside the Landau region, especially along the plasmon dispersion line.

\subsection{The memory function method}

The derivation of the memory function is quite involved so that we describe here only the
different steps necessary in order to arrive to the final result. For a detailed derivation,
we refer the reader to some review articles \cite{a71,a72,a73}. This approach was pioneered by Zwanzig \cite{a70} and Mori \cite{a69}
who built the memory function framework upon Kubo’s non-equilibrium statistical physics.
The starting point is to divide the variables' space into two sub-spaces, that of slowly moving ones,
the so-called conservative quantities, that will affect the macroscopic experimental signal,
and that of fast-moving ones (non-conservative), which are not expected to impact the
experimental signals.  The memory function method includes the effect of the fast
variables into the dynamics of the slow variables. In order to do this, we work with
time-correlation functions or response functions. The connection to the dielectric function
is made through the realization that (i) the dynamical structure factor $S(\mathbf{q},\omega)$ is the
time-Fourier transform of the \emph{classical} density auto-correlation function, (ii) the density-density
response function $\chi_{n n}(\mathbf{q},\omega)$ is related to the dielectric function.

The next step is to show that within this variable space partitioning, auto-correlation functions
satisfy an intregro-differential generalized Langevin equation (GLE)

\begin{equation}\label{eq41}
\frac{d C_A(t)}{dt} \, = \, - \int_0^t \; K_A(t-s) C_A(s) \; ds \quad \text{with} \quad C_A(0) \, = \, 1
\end{equation}

where $C_A(t)$ is the auto-correlation function of interest, and  the unknown function $K_A(t)$
is the so-called memory function which keeps track
of what happened to the system before the present time $t$. The response function $\chi_{A A}(t)$ is related 
to the auto-correlation function through

\begin{equation*}
\chi_{A A}(t) = \frac{d C_A(t)}{dt} \; \left( A | A \right)
\end{equation*}

where $(A,B)$ is the so-called \emph{Kubo scalar product}.

The usual Langevin equation of Brownian
motion corresponds to the Markovian choice $K_A(t) = 1/\tau \, \delta(t)$ where at $t$, the dynamics of the system does
not depend on the previous states of the system. The integral term containing
$K_A(t)$ describes the influence of the fast moving variables, presumably out of reach to the experiment,
on the (conservative) variable $A$ of interest.

It can be demonstrated that the memory function is also an auto-correlation function, and as such it satisfies
a hierarchy of similar GLE

\begin{equation}\label{eq42}
\frac{d {K}_n(t)}{dt} \, = \, 
- \int_0^t \; {K}_{n+1}(t - s) \; {K}_n(s) \; ds
\end{equation}

where we take $C_A(t) = K_0(t)$ and $K_A(t) = K_1(t)$.

The next step is to take the Laplace transform of the integro-differential GLE in order to change it
into an easily solvable equation. This leads to the continued fraction expansion

\begin{equation}\label{eq43}
\bar{C}_{A}(z) \, = \, \frac{K_0(0)}{z + \displaystyle{\frac{K_1(0)}{z + \displaystyle{\frac{K_2(0)}{z +  \displaystyle{\frac{K_3(0)}{\ddots}} } } }}}
\end{equation}

Truncating to order 3 (i.e. expressing the result in terms of 3 moments) and going to the response function representation,  gives

\begin{equation}\label{eq44}
\bar{\chi}_A^{(3)}(z) \, = \,  \displaystyle{
- \frac{ \Delta_0^2 \Delta_1^2 \left[  z + \bar{K}_3(z) \right] }{z^3 + z^2  \bar{K}_3(z) +
z \left(  \Delta_1^2 +  \Delta_2^2 \right) + \Delta_1^2 \;  \bar{K}_3(z)}
} 
\end{equation}

where the $\Delta_n$ are function of the moments $\left\langle\omega^{2n-1}\right\rangle_\chi$ and $K_n(0) = \Delta_n^2$. Here, $\bar{K}_3(z)$ 
is the Laplace transform of the memory function 
$K_3(t)$.

As the moments for $n$ = 0,1,2 are well-known (and to some extent, making use of some approximations, $n = 3$),
the only unknown is the memory function $\bar{K}_3(z)$.

Then, we have, taking $A$ as the (constant) particle density $\bar{n}$

\begin{equation}\label{eq45}
\varepsilon^{-1}(\mathbf{q}, \omega) \, = \, 1 +
\frac{ \omega_p^2 \; \left[ \omega + i \bar{K}_3 (-i  \omega) \right]}{
\omega \; \left[ \omega^2 - \omega_2^2 \right] + i \bar{K}_3 (-i  \omega) \;
\left[ \omega^2 - \omega_1^2 \right]}
\end{equation}
\

We note that under the exchange $Q_2(\omega) \leadsto i \bar{K}_3(-i \omega)$, this equation transforms into the
Nevanlinna equation. The difference is that we have now a physically motivated function $K_3$ for
which many approximations exist and that is related to the relaxation of the system, and hence
in the various decay mechanisms, which could, in principle, be studied experimentally.

\begin{figure}[htp]
\includegraphics[width = 0.9\columnwidth]{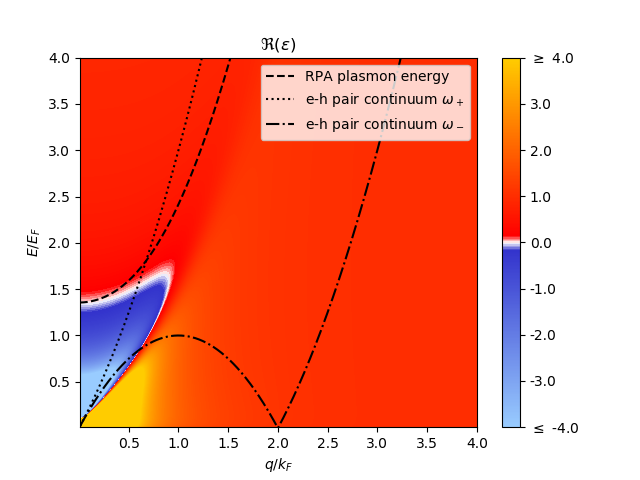}
\includegraphics[width = 0.9\columnwidth]{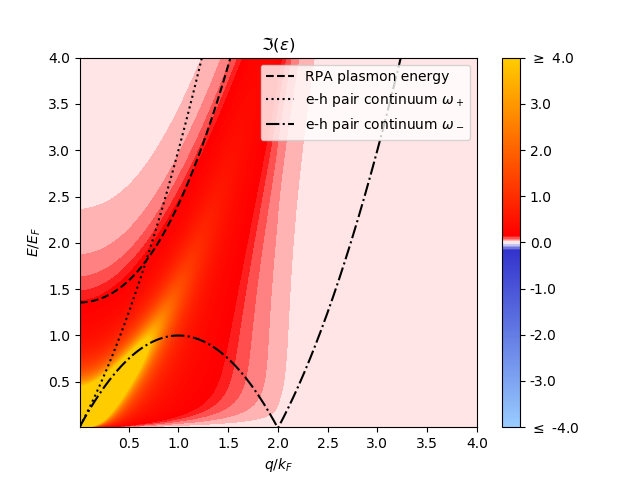}
\caption{Real and imaginary part of the 3-moment memory function dielectric function within the Cole-Cole approximation (COCO) for a value of  0.5 fs.}\label{fig6}
\end{figure}

By definition, through the use of $\left\langle\omega^{1}\right\rangle_\chi$ and $\left\langle\omega^{3}\right\rangle_\chi$, Eq.~\ref{eq45} contains
the conservation of the number of particles and a proper treatment of pair correlation effects.

If we replace the auto-correlation function $A$ in Eq.~\ref{eq43} by a vector containing the density, the
momentum and the energy, the GLE transforms into a matrix equation \cite{a74,a75}. Applying the same approach,
we can derive an expression of the dielectric function that conserves the three previous quantities.
This makes the memory function/memory matrix approach a very powerful and flexible tool to
build up a physically meaningful dielectric function. Moreover, in case of need, two or more memory functions  describing relaxations on different timescales can be used within G\"{o}tze's mode-coupling formalism \cite{a76,a77,a78}. This 
allows to incorporate different decay channels into the modelling of the dielectric function.

Here we consider the Cole-Cole (COCO) type memory function expression ~\cite{a52} for the complex dielectric function describing the relaxation process occurring in the system. This \emph{classic} empirical model can be expressed as ~\cite{a52}
\begin{equation}\label{eq47}
K_{CC}(t) = \tau^2 \dfrac{1}{\Gamma(\alpha -1)}(\frac{t}{\tau})^{\alpha-2} 
\end{equation}
with the power-law exponent $\alpha$ being $0 < \alpha \leq 1 $.  Here, $\Gamma$ is the Gamma function. In the following, we choose $\alpha = 0.5$ and a relaxation time  of $\tau = 0.5$ fs, to be consistent with the results of the Mermin dielectric function.

\begin{figure}[htp]
\includegraphics[width = 0.9\columnwidth]{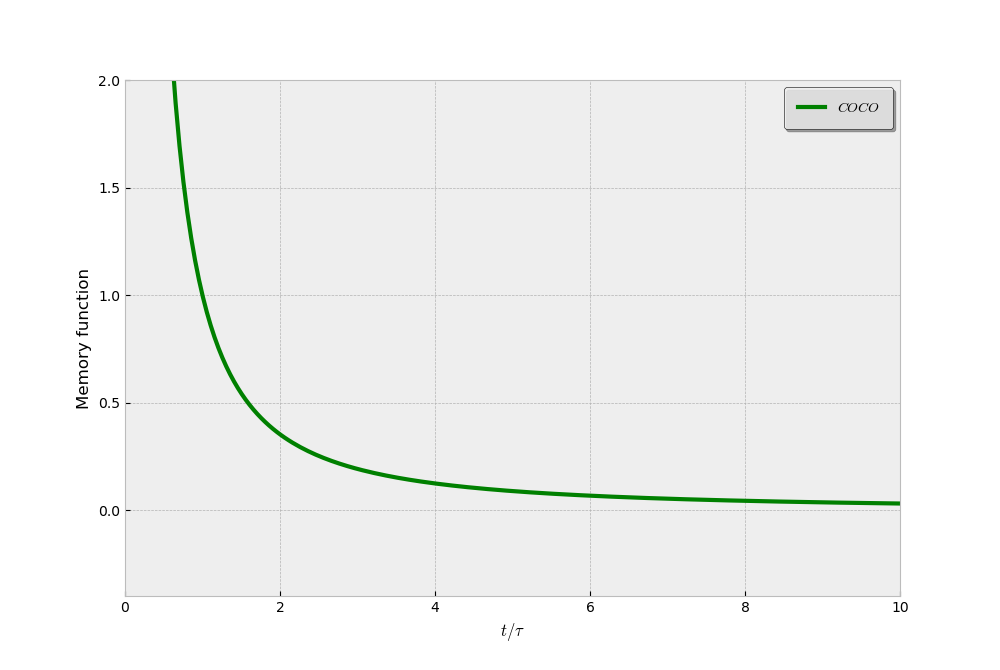}
\caption{The Cole-Cole (COCO) memory function in the time domain.}\label{fig7}
\end{figure}

In Fig. \ref{fig7} we observe the behaviour of memory function in the Cole-Cole approximation at different time-scale. The corresponding dielectric function is represented in Fig. \ref{fig6}. Here, we observe that the imaginary part of the dielectric function also shows some damping of the plasmon outside the Landau region. 

\begin{figure}[htp]
\includegraphics[width = 0.9\columnwidth]{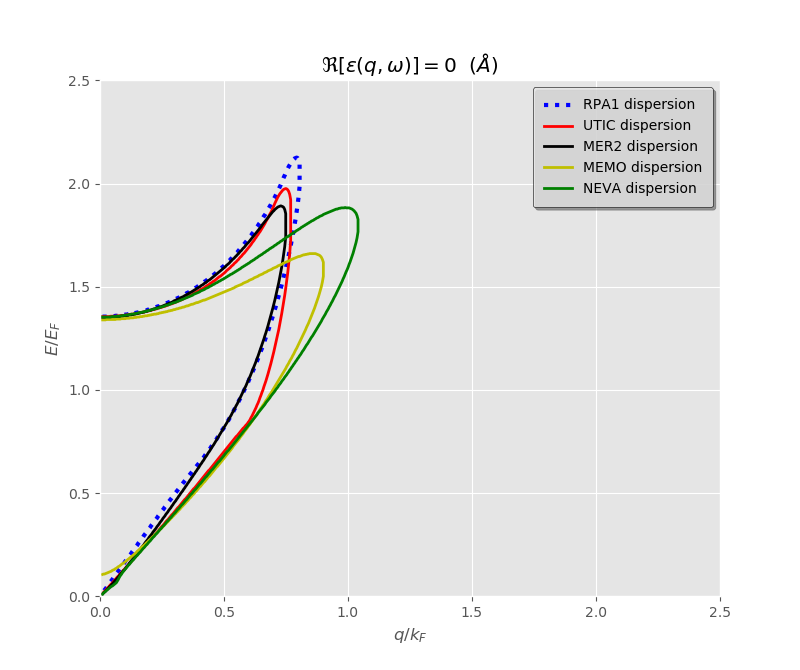}
\caption{Comparison of the collective excitations dispersion bands for different modelling of the dielectric function.}\label{fig8}
\end{figure}

\begin{figure*}[htp]
    \centering
	\subfigure{\includegraphics[width=0.9\columnwidth]{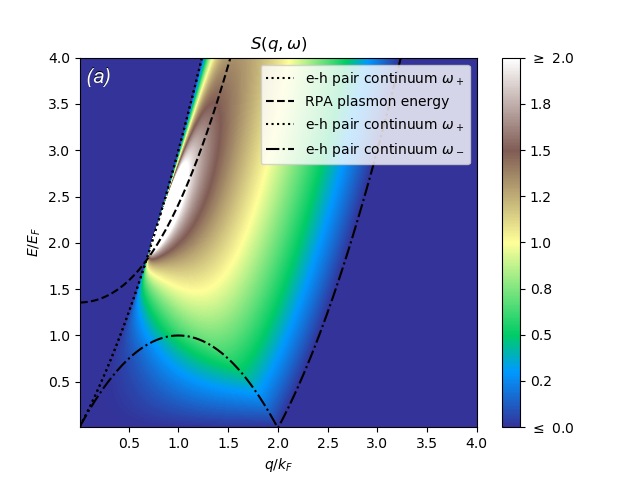}}
    \subfigure{\includegraphics[width=0.9\columnwidth]{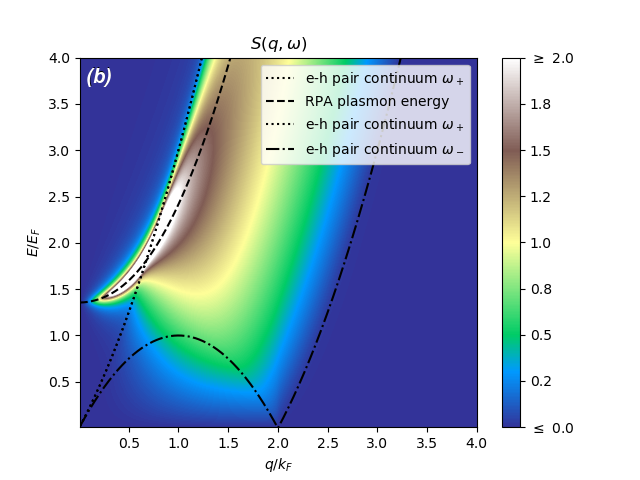}} 
    \subfigure{\includegraphics[width=0.9\columnwidth]{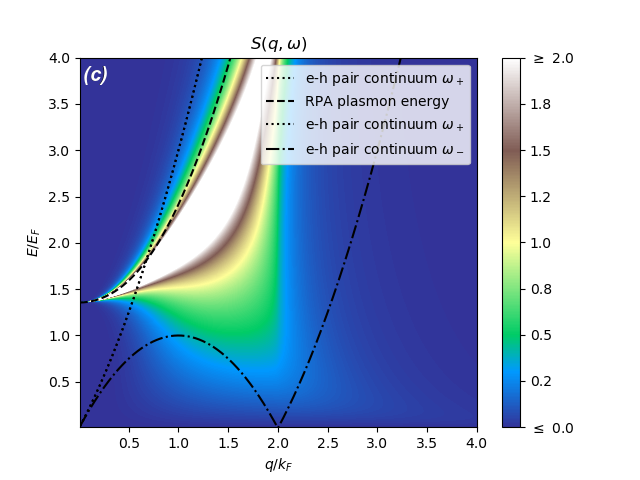}} 
    \subfigure{\includegraphics[width=0.9\columnwidth]{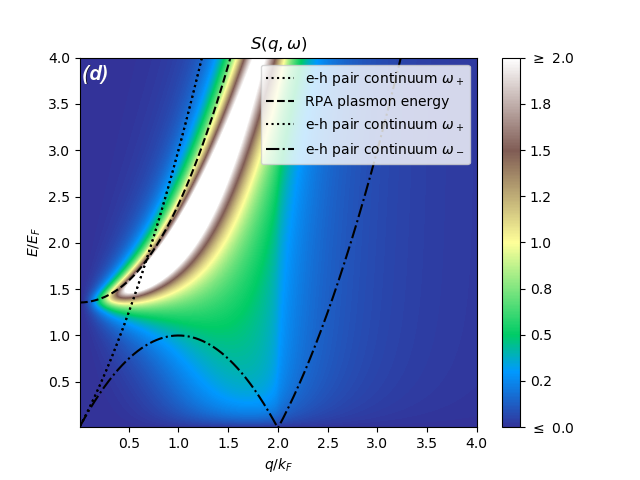}}
    \caption{Respectively (a) RPA, (b) Mermin, (c) Nevanlinna (STA3) and (d) memory function (COCO) structure factors.}\label{fig9}
\end{figure*}

\section{Discussion}

We have explicited in the previous sections different ways to analytically compute model 
dielectric functions, starting from the simple plasmon pole and RPA, and ending up with more involved approaches involving a reconstruction 
in terms of the first moments. We recall that 
our aim in this work is to find a simple (if possible) and flexible method that would describe plasmon dispersion as accurately as possible. Indeed, our ultimate goal being to 
calculate plasmons fluctuation potentials 
in the most general possible case, we need for this to follow Hedin's definition \eqref{eq2}

\begin{equation*}
V^{\mathbf{q}}(\mathbf{r})=\left|\frac{V_{c}(\mathbf{q})}{\displaystyle{\frac{\partial\varepsilon(\mathbf{q},\omega)}{\partial\omega}}\vert_{\omega=\omega_{\mathbf{q}}}}\right|^{1/2}e^{i\mathbf{q}.\mathbf{r}}
\end{equation*}

For low values of $\mathbf{q}$, the fluctuation potential will  be dominated by the Coulomb part. But when $\mathbf{q}$ increases 
(in photoemission, for instance, we will have 
to integrate over $\mathbf{q}$ as the plasmon 
momentum is not detected), the features of the 
first derivative of the dielectric function, taken along the plasmon dispersion will become
more and more important. From this point of view, we see that two features of the dielectric function could become important: 
1) the shape of the plasmon dispersion and 2) a correct description of the plasmon damping. The former corresponds to the upper band of the zeros of the real part of $\varepsilon(\mathbf{q},\omega)$ while the latter is embedded into the imaginary part.

All the calculations were done for a value of the dimensionless Wigner-Seitz radius $r_{s}/a_{0}$=2.079, which corresponds to Aluminium ($r_{s} $= 1.01 \AA).

From figure \ref{fig:foobar}, we see that the plasmon pole does not exhibit any damping and that the plasmon never decays. This means that in principle, we should have to integrate $q$ to infinity. For RPA and correlation-augmented RPA, the integration over $q$ will be limited to a $q_{max}$ above which the plasmon completely decays into an electron-hole pair. This $q_{max}$ is very close 
to the intersection of the plasmon dispersion with the Landau region, which means that in practice no damping will be taken into account as clear from the imaginary part plots. 

The next type of dielectric functions we have considered are the number of electron-conserving Mermin and Hu-O'Connell ones (figure \ref{fig4}). In the former, we used a realistic value of the relaxation time $\tau\,=\, 0.5$ fs. We note that this value is about one order of magnitude larger that the approximation often made in the literature $\tau\, \approx\, 1/\omega_{p}$ which gives here $\tau\, \approx\, 0.04$ fs. This relaxation time approach involves the use of a complex frequency $\bar{\omega} = \omega + i / \tau$ which automatically introduces damping outside the Landau continuum. The related diffusion coefficient method (Hu-O'Connell) also gives such a damping. From this point of view, they are very appealing methods. Unfortunately, being based on the RPA, they lack a proper description of correlation effects. In the case of 
Mermin's approach, this could be remedied for 
by replacing in equation \eqref{eq27} the 
RPA $\varepsilon^0(\mathbf{q},\omega + i/\tau)$ 
by a correlation-augmented one such as RPA + UTI1. 

The final approach we have documented is that based on the 3-moment reconstruction of the dielectric function. This method makes use 
of the moments $\left\langle\omega^{-1}\right\rangle$, 
$\left\langle\omega^{1}\right\rangle$ and 
$\left\langle\omega^{3}\right\rangle$ of 
the loss function. Here, 
$\left\langle\omega^{1}\right\rangle$ ensures 
the conservation of the number of electrons 
while $\left\langle\omega^{3}\right\rangle$ 
incorporates a proper treatment of pair correlations. As can be seen from figure \ref{fig5} and figure \ref{fig6}, the more exact treatment of the correlation changes considerably the shape of the plasmon dispersion with respect to the RPA one (dotted line) or even with respect to RPA + UTI1 (figure \ref{fig:foobar}). In addition, this type of approach does give already some damping of the 
plasmon outside the Landau regime. But it is noteworthy that both calculations have been done 
\emph{without explicit damping}, i.e. using a
 real frequency $\omega$ in equation \eqref{eq39} and \eqref{eq45}, in contrast to the Mermin calculation that was incorporating 
 an explicit damping in the frequency. Therefore, these two methods have even more flexibility that we have used so far. 
 
 A quantity related to the dielectric function that can be  measured experimentally is the dynamical structure factor $S(\mathbf{q},\omega)$. Lemell et al ~\cite{a62} for instance, present in their Fig.2 results for the dynamical structure factor derived from optical data of Mg. 
 There, we seen clearly the plasmon, starting 
 from $q = 0$. In Fig.~\ref{fig9}, we present 
 the structure factors for Aluminium computed 
 from our dielectric function using the (a) RPA, (b) Mermin, (c) Nevalinna-STA3 and (d) memory function-COCO. These plots show that RPA and RPA based dielectric functions should be  ruled out as they are not able to describe properly the plasmon for low-$q$ values, while the other methods give a much better representation that does look like the experimental results of 
Lemell et al ~\cite{a62}.

Now we come to the more general discussion 
of the choice of a suitable dielectric function 
model. We have not explored all the possibilities of the models, but we can already 
outline some important points and perspectives. 
The Mermin approach is limited by being RPA-based, although it could be in principle expanded through the use of LFC. In addition, 
following the scheme devised by  G\"{o}tze \cite{a79} for the computation of the susceptibility, it can be demonstrated that Mermin's dielectric function is a particular case of  
G\"{o}tze's scheme when neglecting 
correlations and using the Markovian form for the memory function $K(\omega)=  1/\tau$.

Nevanlinna functions are hard to find, but as mentioned in the memory function subsection, they are mathematically related to the memory functions through 
$Q_2(\omega) \leadsto i \bar{K}_3(-i \omega)$. Moreover, following equation \eqref{gq}, we see 
that Nevalinna functions and hence memory functions are strongly related to \emph{dynamical} LFC. 

In practice, these considerations mean that most, if not all, the model dielectric functions can be viewed as particular cases of the memory function approach. Another advantage of this method is that it is highly flexible and customizable. We have shown here only the basic features, limiting ourselves to a simple memory function and to a 
scalar generalized Langevin equation, using the constant 
density $\bar{n}$ as the only variable of interest. But, as mentioned in the introduction, by using 
a vector composed of 1) the number density $n_{\mathbf k}(t)$, 2) the longitudinal current density $j^\ell_{\mathbf k}(t)$ and 3) the energy density $e_{\mathbf k}(t)$, we can build a matrix GLE containing a $3\times3$ memory matrix \cite{a74,a75}. Following the same scheme as presented here, we can obtain the expression of a dielectric function that has embedded: (i) the conservation of the number of particle, (ii) the conservation of the momentum, (iii) the conservation of the energy, (iv) pair correlations. Moreover, we believe that the Atwal-Ashcroft approach \cite{a65} will be found to be a particular case of the $3\times3$ memory matrix method.

If need be, this method can be further augmented by building on 4 moments, as it is known that $\left\langle\omega^{5}\right\rangle$ incorporates three-body correlations \cite{a66}. The exact value of this term is very complicated, but reasonable approximations exist that allow to compute it. Work is in progress to incorporate it into the MsSpec-DFM computer code.

The memory function approach we have used here relies 
on a single memory function, which means that it includes 
the relaxation of the system on a single timescale. In classical systems, such as the relaxation of a fluid, molecular dynamics seems to favour two-relaxation-time laws 
\cite{a80}. The memory matrix method we have outlined in the previous paragraph makes use of several memory functions. But even in the simpler method discussed here, we can accommodate several relaxation processes operating on different timescales, which in our case would describe different decay channels of the plasmon. This is the so-called mode-coupling framework \cite{a76,a77,a78} developed by G\"otze.

\section{Conclusion}

In this work, we have tested different electron gas dielectric functions in order to ultimately model fluctuation potentials. 
These fluctuations potentials are the key quantity in the multiple scattering description of plasmon features in 
spectroscopies such as photoemission or EELS.

We have presented three families of dielectric functions, 
RPA-based, damped RPA-based and 3-moment reconstructed.
We have shown that the simple RPA-based ones are not suited 
to our needs. Furthermore, as the damped type of dielectric 
functions can be shown to be a particular case of G\"otze's 
memory function scheme, we have come to the conclusion that 
the memory function approach, to which the Nevalinna one is strongly related, has all the features needed for a precise and accurate modelling. In addition, it can be improved 
by passing to the matrix approach and by incorporating the different plasmon 
decay channels and their timescale through the mode-coupling framework. 

This approach is still based on the homogeneous electron gas model. Therefore, 
no band structure and no crystal structure is involved. We are currently 
investigating the influence of the band structure by performing ab initio 
calculations of $\varepsilon(\mathbf{q},\omega)$ with the Questaal code \cite{a63,a64}.  Preliminary results point towards a minor influence, at least in the case of 
Aluminium which has a simple bandstructure. Another line of research which we are currently pursuing in order to further improve our description is to couple this electron dielectric function to a phonon dielectric function. We are hopeful all 
these improvements will lead to a fast, efficient and flexible approach to 
the modelling of the dielectric function necessary to incorporate plasmons features in to the multiple scattering description of spectroscopies.

The MsSpec-DFM code, built during this work, will be published soon as a separate module of the MsSpec code \cite{a22}. In addition to 3D dielectric functions, as exposed here, it will also contain the modelling of other 
dimensionalities, including graphene-type and multilayer structures.

\section*{Acknowledgements}

A. M. is indebted to Rennes M\'{e}tropole for providing her with two 6-month grants as a visiting scientist.

%\section{Appendix}
%\subsection{Appendix A}

%%%%%%%%%%%%%%%%%%%%%%%%%%%%%%%%%%%%%%%%%%%%%%%%%%%%%%%%%%%%%%%%%%%%%%%%%%%%%%

%%%%%%%%%%%%%%%%%%%%%%%%%%%%%%%%%%%%%%%%%%%%%%%%%%%%%%%%%%%%%%%%%%%%%%%%%%%%%%
%\bibliographystyle{h-physrev4}
\bibliographystyle{apsrev4-1}
%\bibliography{ref1}
%
\end{document}